\documentstyle[12pt]{article}
\newcommand\beq{\begin{equation}}
\newcommand\eeq{\end{equation}}
\begin{document}

\title{\hfill {\rm \normalsize{HD--TVP--93--15}}
\\
\vskip 2cm
Chiral Symmetry Restoration at Finite Temperature in the Linear Sigma--
Model}
\author{D. Metzger, H. Meyer--Ortmanns, H.--J. Pirner\thanks{
Supported by the Bundesministerium
f\"ur Forschung und Technologie (BMFT) under contract 06 HD 729}
\\\\
Institut f\"ur Theoret. Physik,\\Universit\"at Heidelberg
}

\date{ }

\vspace{1.0cm}

\setlength{\baselineskip}{24pt}

\maketitle

\begin{abstract}
\setlength{\baselineskip}{18pt}
\noindent
The temperature behaviour of meson condensates $<\sigma_0>$ and
$<\sigma_8>$
is calculated in the $SU(3)\times SU(3)$-linear sigma model. The couplings
of the Lagrangian are fitted to the physical $\pi,K,\eta,\eta'$ masses,
the pion decay constant and a $O^+(I=0)$ scalar mass of $m_\sigma=1.5$
GeV.
The quartic terms of the mesonic interaction are converted to a quadratic
term with
the help of a Hubbard-Stratonovich transformation. Effective mass terms are
generated this way, which are treated self-consistently to leading order of a
$1/N$-expansion. We calculate the light $<\bar q q>$ and strange $<\bar s
s>$-quark
condensates using PCAC relations between the meson masses and
condensates.
For a cut-off value of 1.5 GeV we find a first-order chiral transition at a
critical
temperature $T_c\sim 161$ MeV. At this temperature the spontaneously
broken subgroup
$SU(2)\times SU(2)$ is restored. Entropy density, energy density and
pressure are
calculated for temperatures up to and slightly above the critical
temperature. To our
surprise we find some indications for a reduced contribution from strange
mesons for
$T\geq T_c$.
\noindent
\end{abstract}
\newpage
\section{Introduction}
The study of finite temperature QCD is important both for theoretical and
experimental hadron physics. Theoretically  a hadron gas in thermal
equilibrium is
the  simplest system to study the dynamics underlying deconfinement and
chiral
symmetry restoration. Experimentally a nearly equilibrated hadron gas with
a
transverse radius $R_\bot \approx 10$~fm is supposed to give the dominant
contribution at a later stage of relativistic heavy ion collisions. In
principle
this allows an experimental check of the equation of state of hadrons. In the
following work we approach the phase transition region from the low
temperature
side. Below $T\sim 100$~MeV  pions are known to be the most relevant
degrees of
freedom. Above this temperature region also heavier hadrons give a
non-negligible contribution to the condensates and thermodynamic
quantities
\cite{ger}.  One way of including part of the heavier mesons is provided by
the choice
of $SU(3)\times SU(3)$ as chiral symmetry group rather than
$SU(2)\times SU(2)$. The linear
$SU(3)\times SU(3)$
sigma model includes a nonet of pseudoscalar  $(O^-)$-fields and a nonet of
scalar $(O^+)$-fields \cite{gas}. The spontaneous breaking of the
$SU(3)\times SU(3)$ symmetry leads to
massless $(O^-)$ Goldstone modes. Obviously a massless pseudoscalar octet
does
not provide an adequate approximation to the experimentally observed
meson spectrum.
Therefore we include explicit symmetry breaking terms to account for the
physical
mass values of the octet-fields. A cubic term in the meson fields guarantees
the correct mass splitting of the $\eta-\eta'$ masses which is due to the
$U(1)$-anomaly. It reflects  the 't Hooft-determinant on the quark level.

\section{The model at zero temperature}
For a Euclidean metric the Lagrangian of the linear sigma--model is given
as
\newpage
\begin{eqnarray}
\label{eq:1}
{\cal L} & = & \frac{1}{2}\partial_{\mu}\Phi\partial_{\mu}\Phi^+
-\frac{1}{2}\mu_{0}^{2} {\rm tr}\Phi\Phi^++f_{1}\left(
{\rm tr}\Phi\Phi^+
\right)^{2}+f_{2}{\rm tr}\left(\Phi\Phi
^+\right)^2\nonumber\\
 & & +g\left(\det\Phi+\det\Phi^+\right)+\varepsilon_{0}\sigma_{0}+
\varepsilon_8\sigma_{8},
\end{eqnarray}
where the ($3\times 3$) matrix field $\Phi(x)$ is given in terms of  Gell--
Mann
matrices $\lambda_\ell\  (\ell=0,\ldots ,8$) as

\begin{equation}
\Phi=\frac{1}{\sqrt{2}}\sum_{\ell=0}^{8}\left(\sigma_{\ell}+i\pi_{\ell}
\right)\lambda_{\ell}.
\end{equation}
Here $\sigma_{\ell}$ and $\pi_{\ell}$ denote the nonets of scalar and
pseudoscalar
mesons, respectively. As order parameters for the chiral transition
we choose the meson condensates $<\sigma_0>$ and $<\sigma_8>$. The chiral
symmetry of
${\cal L}$  is explicitly broken by the term
$(-\varepsilon_0\sigma_0-\varepsilon_8\sigma_8)$, corresponding to the
finite quark
mass term $m_q\bar q q+m_s \bar s s$ on the quark level. The chiral limit is
realized
for vanishing external fields $\varepsilon_0$ and $\varepsilon_8$. Note that
the
action $S=\int d^3x d\tau{\cal L}$ with ${\cal L}$ of Eq.~(\ref{eq:1}) may be
regarded as an effective action for QCD, constructed in terms of an order
parameter
field $\Phi$ for the chiral transition. It plays a similar role to Landau's
free
energy functional for a scalar  order parameter field for investigating the
phase
structure of a $\Phi^4$-theory. First conjectures about the chiral phase
transition
were based on a renormalization group analysis in momentum space
\cite{pis,iac}. An
$\varepsilon$-expansion in $d=4-\varepsilon$ dimensions has been
performed by
Iacobson and Amit \cite{iac}. When it is applied to the Lagrangian of
Eq.~(\ref{eq:1})
with $g=0,\varepsilon_0=0=\varepsilon_8$, it predicts a first order chiral
transition.
For three flavours the det-term is cubic in the field components. Hence a
non-vanishing $g$ will further support the first order nature of the
transition. In
contrast finite mass terms $(\varepsilon_0\not= 0\not=\varepsilon_8)$ may
change the
transition to a smooth crossover behaviour, if their values are large enough.

Patk\'os  and Frei confirmed the first order nature of the chiral transition in
the
chiral limit of the
$SU(3)\times SU(3)$ linear sigma model \cite{frei}. The result was obtained
in a saddle
point approximation to the free energy functional in three dimensions
(dropping the
imaginary time-dependence of the fields $\Phi$ in Eq.~(\ref{eq:1})). In a
subsequent
work \cite{mey} it was shown that non-vanishing pseudoscalar  meson
masses actually
change the first oder transition to a smooth crossover in the condensates, if
otherwise the same approach is followed as in \cite{frei}.

In the present work we have extended the former approach to a treatment
of the full
four-dimensional theory, keeping all Matsubara frequencies in the effective
potential. The reason is that a complete dimensional reduction from 4 to 3
dimensions
may be approximately realized at high temperatures or for a second order
phase
transition.  In the present work we are interested in the low  temperature
region; we also cannot expect a second order transition in the presence of
explicit
symmetry breaking terms. Therefore contributions from non-zero Matsubara
frequencies
may even qualitatively change the results. This is in fact what  we will
demonstrate in
this paper.

The six unknown couplings of the sigma-model (Eq.~(\ref{eq:1}))\
$(\mu^2_0, f_1, f_2,
g,\varepsilon_0,\varepsilon_8)$ are assumed to be temperature independent
and fitted
to the pseudoscalar masses  at zero temperature. Further experimental input
parameters
are the pion decay constant  $f_\pi=93$ MeV and a high lying $(O^+)$  scalar
mass
$m_\sigma=1.59$ GeV (cf. Table 1). For the remaining scalar masses and the
coupling
constants we obtain the values of Table 1.

\begin{table}[h]
\begin{center}
\begin{tabular}{|l|l|l|l|l|l|}
\hline
\multicolumn{6}{|c|}{Input}\\
\hline\hline
$m_{\pi}$ [MeV]& $m_{K} [MeV] $ & $m_{\eta} [MeV]$ & $m_{\eta'}$
[MeV]&
$f_{\pi}$ [MeV] & $m_{\sigma}=m_{f_{0}}$ [MeV]\\
\hline
138.04 & 495.66 & 547.45 & 957.75 & 93 & 1590\\
\hline\hline
\multicolumn{6}{|c|}{Output}\\
\hline\hline
$\mu_{0}^{2}$ [GeV$^{2}$] & $f_{1}$ & $f_{2}$ & $g$ [GeV] &
$\varepsilon_{0}$
[GeV$^{3}$] & $\varepsilon_{8}$ [GeV$^{3}$]\\
\hline
0.758 & 12.166 & 3.053 & 1.527 & 0.02656 & -0.03449\\
\hline\hline
$m_{a_{0}}$ [MeV] & $m_{K_{0}^{\ast}}$ [MeV] & $m_{f_{0}'}$ [MeV] &
$f_{K}$
[MeV] & & \\
\hline
914.05 & 913.35 & 764.71 & 128.81 & & \\
\hline
\end{tabular}
\caption{\label{values}Tree level parametrization of the $SU(3)\times SU(3)$
linear
sigma model (input data taken from experiment).} \end{center}
\end{table}
The interpretation of the observed scalar mesons is controversial. There are
good reasons to interpret the ($0^+$) mesons at 980 MeV
as meson bound states~\cite{wei}. The model underestimates the strange
quark mass
splitting in the scalar meson sector, the value for $m_{K^*_0}$ comes out too
small.

The effective theory can be related to the underlying QCD Lagrangian by
comparing
the symmetry breaking terms in both Lagrangians and identifying terms
with the same
transformation behaviour under
$SU(3)\times SU(3)$. Taking expectation values in these equations
we obtain the following relations between the light quark condensates,
strange quark
condensates and meson condensates

\begin{eqnarray}
\label{eq:3}
<\bar q q>&=&\frac{(-\varepsilon_0)}{2\hat m+m_s}\ <\sigma_0>+\frac{(-
\varepsilon_8)}
{2(\hat m-m_s)}<\sigma_8>\nonumber\\
<\bar s s>&=&\frac{(-\varepsilon_0)}{2\hat m+m_s}\ <\sigma_0>-\frac{(-
\varepsilon_8)}
{(\hat m-m_s)}<\sigma_8>.
\end{eqnarray}
We use $\hat m\equiv (m_u +m_d)/2
=(11.25\pm 1.45)$ MeV and $m_s=(205\pm 50)$
MeV for the light and strange quark masses at a scale $\Lambda=1$
GeV~\cite{nar}.  From
the scalar meson condensates at $T=0$,  $\left<\sigma_{0}\right>=0.144$
GeV and $\left<\sigma_{8}\right>=-0.0415$ GeV we get

\begin{eqnarray}
\left<\bar{q}q\right> & = & -\left(235\pm 60 {\rm
MeV}\right)^{3}\nonumber\\
\left<\bar{s}s\right> & = & -\left(290\pm 30 {\rm MeV}\right)^{3}
\end{eqnarray}
in accordance with values  from PCAC relations~\cite{nar} within the
error bars. Since we treat the coefficients $\varepsilon_0,\varepsilon_8$ of
$<\sigma_0>$ and $<\sigma_8>$, and $\hat m, m_s$ of $<\bar q q>$ and
$<\bar s s>$
as temperature independent, we will use Eqs.~(\ref{eq:3}) for all
temperatures
to translate our results for meson condensates into quark condensates.

We also check that the pseudoscalar meson mass squares, in particular
$m^2_\pi$ and
$m^2_K$ are linear functions of the symmetry breaking parameters
$\varepsilon_0,\varepsilon_8$. Varying $\varepsilon_0,\varepsilon_8$
while keeping
the other couplings fixed we can simulate the sigma model at unphysical
meson
masses. Since the current quark masses are assumed to depend  linearly on
$\varepsilon_0$ and $\varepsilon_8$, an arbitrary meson  mass set can be
related to
a mass point in the $(m_{u,d},m_s)$-plane by specifying the choice of
$(\varepsilon_0,\varepsilon_8)$. This may be useful in order to compare our
results
for meson (and quark) condensates with lattice simulations of the chiral
transition.

\section{Thermodynamics}
The thermodynamics of the linear sigma model is determined by the
partition
function with the Lagrangian of Eq.~(\ref{eq:1})

\begin{equation}     \label{eq:5}
Z=\int{\cal D}\Phi \exp\left\lbrace-\int^\beta_0 d\tau\int d^3 x{\cal L}
(\Phi(\vec x, \tau))\right\rbrace.
\end{equation}
We will treat $Z$ in a saddle point approximation. The saddle point
approximation
amounts to the leading order of a $1/N$-expansion in this model, where
$N=2 N^2_f=18$. Note that $\cal L$ of Eq.~(\ref{eq:1}) would be $O(N)$-
invariant, if
$f_2=0$ and $g=0$. Our input parameters lead to non-vanishing values
of $f_2$ and $g$, therefore the $O(N)$-symmetry is only approximately
realized.

We calculate the effective potential as a constrained free energy density
$U_{eff}
(\xi_0,\xi_8)$, that is the free energy density of the system under the
constraint
that the average values of $\sigma_0$ and $\sigma_8$ take some prescribed
values
$\xi_0$ and $\xi_8$.  The values $\xi_{0_{min}}$ and $\xi_{8_{min}}$ that
minimize $U_{eff}$, give the physically relevant, temperature dependent
vacuum
expectation values, i.e. $<\sigma_0>=\xi_{0_{min}},\
<\sigma_8>=\xi_{8_{min}}$.
Hence we start with the background field ansatz
\begin{eqnarray}\label{eq:6}
\sigma_0&=& \xi_0+\sigma'_0\nonumber\\
\sigma_8&=& \xi_8+\sigma'_8,
\end{eqnarray}
where $\sigma'_0$ and $\sigma'_8$ denote the fluctuations around the
classical
background fields $\xi_0$ and $\xi_8$. All other field components are
assumed to have
zero vacuum expectation value, i.e. $\sigma_\ell=\sigma'_\ell$ for
$\ell=1,\ldots,7$
and $\pi_\ell=\pi'_\ell$ for $\ell=0,\ldots,8$.
The relation between the effective potential $U_{eff}$ and $Z$ is given by
\newpage
\begin{eqnarray}
\label{eq:7}
Z&=&\int d\xi_0\int d\xi_8\hat Z(\xi_0, \xi_8)\nonumber\\
\hat Z(\xi_0,\xi_8)&=&: e^{-\beta V U_{eff}(\xi_0,\xi_8)}\nonumber\\
&=&\int {\cal D}\Phi\delta \left[\int\sigma_0(\vec
x,\tau)-\xi_0\right]\delta\left[\int\sigma_8(\vec x,\tau)-
\xi_8\right]\nonumber\\
& &\cdot\prod_{\ell\not=0,8}\delta\left[\int\sigma_\ell(\vec x,\tau)\right]
\prod^8_{\ell=0}\delta\left[\int\pi_\ell(\vec x,\tau)\right]\cdot e^{-
\int^\beta_0
d\tau\int d^3 x{\cal L}[\Phi]}
\end{eqnarray}
where $'\int'$ is a short-hand notation for $\frac{1}{\beta V}\int^\beta_0
d\tau\int
d^3 x$.

Next we insert the background field ansatz (\ref{eq:6}) in $\cal L$ and
expand the
Lagrangian in powers of $\Phi'=\frac{1}{\sqrt 2}\sum^8_{\ell=0}(\sigma'_\ell
+i\pi'_\ell)\lambda_\ell$. The constant terms in $\Phi'$ lead to the classical
part
of the effective potential $U_{class}$

\begin{eqnarray}
\label{eq:8}
U_{class}(\xi_{0},\xi_{8}) & = &
-
\frac{1}{2}\mu_0^2\left(\xi_{0}^{2}+\xi_{8}^{2}\right)+\frac{g}{3\sqrt{3}}\cdot
\left(2\xi_{0}^{3}-\sqrt{2}\xi_{8}^{3}-3\xi_{0}\xi_{8}^{2}\right)
-\frac{2\sqrt{2}}{3}f_{2}\xi_{0}\xi_{8}^{3}\\
& &+\left(f_{1}+\frac{f_{2}}{3}
\right)\xi_{0}^{4}+\left(f_{1}+\frac{f_{2}}{2}\right)\xi_{8}^{4}
+2\left(f_{1}+f_{2}\right)\xi_{0}^{2}\xi_{8}^{2}-\varepsilon_{0}\xi_{0}
-\varepsilon_{8}\xi_{8}.\nonumber
\end{eqnarray}

Linear terms in $\Phi'_\ell$ vanish for all $\ell=0,\ldots,8$ due to the
$\delta$-constraints in Eq.~(\ref{eq:7}).

Quadratic terms in $\Phi'$ define the isospin multiplet masses $m^2_Q$,
where
$Q=1,\ldots,8$ labels the multiplets. The contribution to the Lagrangian is
denoted
by ${\cal L}^{(2)}$
\begin{equation} \label{eq:9}
{\cal L}^{(2)}=\frac{1}{2}\sum_Q
g(Q)\left(\partial_\mu\varphi'_Q\partial_\mu
\varphi^{\prime
\dagger}_Q+m^2_Q\varphi'_Q\varphi^{\prime\dagger}_Q\right).
\end{equation}
Here $\varphi'_Q$ denotes $\sigma'_Q$ for $Q=1,\ldots,4$ and $\pi'_Q$
for $Q=5,\ldots,8,\ g(Q)$ is the multiplicity of the isospin multiplet. We have
$g(1)=3$ for the pions, $g(2)=4$ for the kaons, $g(3)=1=g(4)$ for $\eta,\eta'$,
respectively. Correspondingly, the multiplicities for the scalar nonets are
$g(5)=3,
\ g(6)=4,\ g(7)=1,\ g(8)=1$ for the $a_0,K^*_0, f_0,f'_0$-mesons. Typical
expressions for the masses are the pseudoscalar and scalar triplet masses
\newpage
\begin{eqnarray}
m_{\pi}^{2} & = &
-\mu_{0}^{2}+\left(4f_{1}+\frac{4}{3}f_{2}\right)\xi_{0}^{2}+\left(4f_{
1}+\frac{2}{3}f_{2}\right)\xi_{8}^{2}\nonumber\\
& &+\frac{4}{3}\sqrt{2}f_{2}\xi_{0}
\xi_{8}
  +\frac{2}{\sqrt{3}}g\xi_{0}-2\sqrt{\frac{2}{3}}g\xi_{8}\label{eq:10}\\
m_{a_{0}}^{2} & = &
-\mu_{0}^{2}+\left(4f_{1}+4f_{2}\right)\xi_{0}^{2}+\left(4f_{
1}+2f_{2}\right)\xi_{8}^{2}\nonumber\\
& &+\frac{8}{\sqrt{2}}f_{2}\xi_{0}\xi_{8}
-\frac{2}{\sqrt{3}}g\xi_{0}+2\sqrt{\frac{2}{3}}g\xi_{8}.\label{eq:11}
\end{eqnarray}

The cubic part in $\Phi'$ will be neglected, while the quartic term ${\cal
L}^{(4)}$
\begin{equation}\label{eq:12}
{\cal L}^{(4)}=f_1({\rm tr} \Phi'\Phi^{\prime\dagger})^2+f_2 {\rm tr}
(\Phi'\Phi^{\prime\dagger})^2
\end{equation}

is quadratized by introducing an auxiliary matrix field $\sum(\vec x,\tau)$.
This is a matrix version of a Hubbard-Stratonovich transformation
\cite{hub}. We have
the identity

\begin{eqnarray}
\label{eq:13}
& &\exp\left\lbrace \int_{0}^{\beta} d\tau\int
d^{3}x (-)\left[ f_{1}\left( {\rm tr}\,\Phi'\Phi^{\prime\dagger}\right)
^{2}+
f_{2}{\rm tr}\,\left( \Phi'\Phi^{\prime\dagger}\right) ^{2}\right]
\right\rbrace
\nonumber\\
& = & {\rm const}\cdot\int_{c-i\infty}^{c+i\infty}{\cal D}\Sigma\:
\exp\left[\frac{1}{16(\varepsilon+3\alpha)^{2}}\int_{0}^{\beta}\right.
d\tau\int d^{3}x\nonumber\\
& &\left\{ {\rm
tr}\,\Sigma^{2}+2\mu^2_0 {\rm tr}\right. \Sigma
 -8(\varepsilon+3\alpha)[\varepsilon{\rm
tr}\,\left(\Sigma\Phi'\Phi^{\prime\dagger}\right)+\nonumber\\
& &+\alpha{\rm tr}\,\Sigma\cdot{\rm
tr}\,\left(\Phi'\Phi^{\prime\dagger}\right)+\mu_{0}^{2}\varepsilon{\rm tr}
(\Phi'\Phi^{\prime\dagger})+3\alpha\mu^2_0{\rm tr}\left.\left.
(\Phi'\Phi^{\prime\dagger})]\right\}
\right],
\end{eqnarray}
where
\begin{eqnarray}
 2\varepsilon\alpha+3\alpha^2&\equiv& f_1\nonumber\\
\varepsilon^2&\equiv &f_2\quad.
\end{eqnarray}

In the saddle point approximation we drop $\int{\cal D}\Sigma$. As $SU(3)$-
symmetric
ansatz we use a constant diagonal matrix

\begin{equation}
\label{eq:14}
\Sigma=\left(\begin{array}{ccc}
                                 s & 0 & 0\\
                                 0 & s & 0\\
                                 0 & 0 & s
                            \end{array}
\right)\quad.
\end{equation}

The choice of $s$ will be optimized later, cf. Eq.~(\ref{eq:27}). When $Z$
is rewritten upon using Eq.~(\ref{eq:13}), ${\cal L}^{(4)}$ of
Eq.~(\ref{eq:12})
is replaced  by ${\cal L}^{(4)\prime}$ given as

\begin{equation}
\label{eq:15}
{\cal
L}^{(4)\prime}=-
\frac{3}{8(3f_{1}+f_{2})}\left(\frac{s^{2}}{2}+\mu_{0}^{2}s\right)
+\frac{1}{2}\left(s+\mu_{0}^{2}\right)\,{\rm
tr}\,(\Phi'\Phi^{\prime\dagger}).
\end{equation}

Hence the effect of the quadratization procedure is to induce an extra mass
term
$(s+\mu^2_0)$ and a contribution $U_{\rm saddle}$ to $U_{eff}$, which is
independent
of $\xi_0$ and $\xi_8$.

\begin{equation}
\label{eq:16}
U_{\rm saddle} =  -\frac{3}{8(3f_{1}+f_{2})}\left(\frac{s^{2}}{2}+\mu_{0}^
{2}s\right).
\end{equation}

We are not aware of an analogous
identity to Eq.~(\ref{eq:13}), which includes the det-term of $\cal L$ and
leads to a
tractable form. Therefore  we drop the cubic  term in $\Phi'$ as mentioned
above.

This way we finally end up with the following expression for $\hat Z$

\begin{eqnarray}
\label{eq:17}
\hat Z(\xi_0,\xi_8)&=&e^{-\beta V(U_{\rm class}+U_{\rm
saddle})}\cdot\nonumber\\
& &\cdot\int\prod^8_{Q=1}{\cal D}\varphi'_Q e^{-\int^\beta_0
d\tau\int d^3 x\frac{1}{2}\sum_Q g (Q)(\partial_\mu\varphi'_Q\partial_\mu
\varphi^{\prime\dagger}_Q+X^2_Q\varphi'_Q\varphi^{\prime\dagger}_Q)}
\end{eqnarray}
where
\begin{equation}\label{eq:18}
X^2_Q\equiv m^2_Q+\mu^2_0+s.
\end{equation}

Thus we are left with an effectively free field theory. The only remnant of
the
interaction appears in the effective mass squared $X^2_Q$ via the auxiliary
field
$s$.

The choice of a self-consistent effective meson  mass squared has been
pursued
already in Refs.~\cite{frei,mey}.  This is an essentially new ingredient
compared to
earlier calculations of the chiral transition in the linear sigma model
\cite{gol}.
The positive contribution of $s$ to the effective mass extends the
temperature region,
where imaginary parts in the effective potential can be avoided. In general,
imaginary parts are encountered, when the effective mass-arguments of
logarithmic
terms become negative. They are an artifact of the perturbative evaluation
of the
effective potential and of no physical significance, as long as the volume is
infinite.
In our application the optimized choice for $s$ will increase as function of
temperature and lead to positive $X^2_Q$ over a wide range of parameters.

Gaussian integration over the fluctuating fields $\Phi'$ in Eq.~(\ref{eq:17})
gives
\begin{eqnarray}\label{eq:19}
\hat Z(\xi_0,\xi_8)&=&\exp\left\lbrace-\beta V\left[ U_{\rm class}+U_{\rm
saddle}+
\right.\right.\nonumber\\
& & +\frac{1}{2\beta}\sum^8_{Q=1}g(Q)\sum_{n\in
Z\!\!\!\!/}\int\frac{d^3k}{(2\pi)^3}
\left.\left.\ln\left(\beta^2(\omega^2_n+\omega^2_Q)\right)\right]\right\rbrace
\end{eqnarray}
where
\begin{equation}\label{eq:20}
\omega^2_Q\equiv k^2+X^2_Q,
\end{equation}
and
\begin{equation}
\label{eq:21}
\omega^2_n\equiv(2\pi n/T)^2
\end{equation}
denote the Matsubara frequencies. In contrast to our former approach
\cite{mey}
we keep all Matsubara frequencies and evaluate $\sum_{n\in Z\!\!\!\!/}$ in
the
standard way, see e.g. \cite{kap}. The result is

\begin{eqnarray}
\hat Z(\xi_0,\xi_8;s)&=&e^{-\beta V U_{\rm
eff}(\xi_0,\xi_8;s)}\label{eq:22}\\
U_{\rm eff}(\xi_0,\xi_8;s)&=& U_{\rm class}+U_{\rm saddle} +U_{\rm th}
+U_0
\label{eq:23}\\
U_{\rm th} & \equiv & \frac{1}{\beta}\sum_{Q=1}^{8}g(Q)\int
\frac{d^{3}k}
{(2\pi)^{3}}\ln\left(1-e^{ -\beta\omega_Q}\right)\label{eq:24}\\
U_{0} & \equiv & \frac{1}{2}
\sum_{Q=1}^{8}g(Q)\int^\Lambda\frac{ d^{3}k}
{(2\pi)^{3}}\omega_Q.\label{eq:25}
\end{eqnarray}

Here we have indicated that $\hat Z$ and $U_{\rm eff}$
still depend explicitly on the auxiliary field $s$. The integral in
Eq.~(\ref{eq:25}) is regularized with a three-momentum cut-off
$\Lambda$. The thermal contribution $U_{\rm th}$  vanishes at zero
temperature and is
finite for $T>0$, while the zero point energy $U_0$ diverges as
$\Lambda\to\infty$.

The linear sigma model is a renormalizable theory, and the cut-off  could be
removed
after a suitable renormalization prescription. Since we are dealing with an
effective
model, the need for such a renormalization may be less obvious. Anyway we
do not
believe in this model as an effective description for QCD, when the momenta
exceed a
certain scale, say $\Lambda\approx 1 - 1.5 $ GeV. The necessity for a
renormalization
arises, when we postulate a matching between the physical masses and
condensates with
the  $T=0$-values, and $T$ approaches zero  from above. Such a matching is
guaranteed, if we impose the following subtractions on the zero point energy
part

\begin{eqnarray}
\label{eq:26}
& &U_{0}^{\rm ren}(X^2_Q(\xi_0,\xi_8);\Lambda):  =  U_{0}
(X^2_Q)-\lbrace U_0(m^2_{\rm phys})+\nonumber\\
& & \frac{\partial U_0(m^2)}{\partial m^2}|_{m^2_{\rm phys}}
\cdot(X^2_Q-m^2_{\rm phys})+\frac{1}{2}\frac{\partial^2 U_0(m^2)}
{\partial(m^2)^2}
|_{m^2_{\rm phys}}\cdot(X^2_Q-m^2_{\rm phys})^2\rbrace.
\end{eqnarray}

\vspace{0.5cm}
Here $m^2_{\rm phys}$ is given  by $m^2_Q$ of Eq.~(\ref{eq:9}) evaluated at
$\xi_0=<\xi_0>$ and $\xi_8=<\xi_8>$,  i.e. for physical condensate values. The
optimal choice $s^*$ for the auxiliary field $s$ is then determined by
\begin{equation}\label{eq:27}
\frac{\partial U^{\rm ren}_{\rm eff}}{\partial s}|_{s^*}=0,
\end{equation}
where $U^{\rm ren}_{\rm eff}$ equals $U_{\rm eff}$ of Eq.~(\ref{eq:23})
with
$U_0$ replaced by  $U_0^{\rm ren}$ of Eq.~(\ref{eq:26}).

Upon using Eq.~(\ref{eq:27}) it is easily verified that $<\xi_i>$, defined as
\begin{equation}\label{eq:28}
<\xi_i>=\frac{1}{\beta V}\frac{\partial\ln Z}{\partial \varepsilon_i},\quad
i=0,8,
\end{equation}
is free  of extra contributions from the zero point energies at $T=0$ if $\ln
Z=-\beta VU^{\rm ren}_{\rm eff}
(\xi_0, \xi_8;s^*)$.  Thus a matching with $<\xi_i>_T=0$ is ensured. Similarly
we find for the effective masses
\begin{equation}\label{eq:29}
X^2_Q|_{T=0}=m^2_Q+s+\mu^2_0=m^2_{\rm phys}
\end{equation}
for $\xi_0=<\xi_0>,\ \xi_8=<\xi_8>,$ since $s=-\mu^2_0$ at $T=0$.

Note that the sensitivity to the cut-off in Eq.~(\ref{eq:26}) is reduced from a
$\Lambda^4$-dependence to a $1/\Lambda^2$-dependence. This is a
desirable feature in
view of the uncertainties in a suitable choice for $\Lambda$.
We have taken $\Lambda=1.5$ GeV and kept the cut-off finite throughout
the
calculations.

A further argument in favour of keeping the cut-off finite relies on results of
Bardeen and Moshe \cite{bar} on the $1/N$-expansion in $O(N)$-theories.
According to
these results a symmetry restored groundstate occurs as vacuum  state at
zero
temperature, if the cut-off is sent to infinity.

Now we are prepared to determine the temperature dependence of the
order parameters
$<\xi_0>(T),\ <\xi_8>(T)$  from the minima of $U^{\rm ren}_{\rm eff}
(\xi_0,\xi_8;
s^*)$. Thermodynamic quantities like energy densities, entropy densities and
pressure
can be derived from $Z$ in the standard way, if $Z$ is approximated as
\begin{equation}\label{eq:30}
\hat Z^{\rm ren}\equiv e^{-\beta V U^{\rm ren}_{\rm eff}(\xi_0,\xi_8;s^*)}.
\end{equation}

\section{Results}
For the parameters of Table~\ref{values} we vary the temperature and
determine
for each $T$ the extremum of $U_{\rm eff}$ as a function of $\xi_{0}$,
$\xi_{8}$ and $s$. The extremum is a minimum with respect to
$\xi_{0}$ and $\xi_{8}$ and a maximum with
respect to  $s$. (For zero temperature this is easily seen from the explicit
form
of $U_{\rm class}$ (Eq.~(\ref{eq:8})) and $U_{\rm saddle}$
(Eq.~(\ref{eq:16})).
Numerically the most convenient procedure is to determine the common
zeroes in the
derivatives $\frac{\partial U_{\rm eff}}{\partial \xi_0},\ \frac{\partial
U_{\rm eff}}
{\partial \xi_8}$ and $\frac{\partial U_{\rm eff}}{\partial s}$ first, and then
to
check the desired minimum/maximum properties of the saddle point.

In Figs. 1 and 2 we show the variations of $\frac{<\bar q q>(T)}{<\bar q
q>T=0}$
and $\frac{<\bar s s>(T)}{<\bar s s>T=0}$ as a function of temperature
obtained from
$<\xi_0>(T)$ and $<\xi_8>(T)$ with the help of Eq.~(\ref{eq:3}). We observe a
gradual decrease of the light quark condensate, whereas the strange quark
condensate
stays remarkably constant.  A first order transition occurs at $T_c=$ 161
MeV for
$\Lambda=$ 1.5 GeV.
Note that we have still a cubic term in the classical part of the potential
(Eq.~(\ref{eq:8})), the term we have dropped is cubic in the fluctuating
fields.
The
critical temperature is determined in such a way that the pressure is
continuous at
$T_c$ as a function of temperature. The values for the mesonic condensates
show
pronounced hysteresis effects, which are characteristic for a first  order
phase
transition.

At $T_c$ the strange quark condensate does not drop to zero. Only the
$SU(2)\times
SU(2)$ part of the chiral symmetry is restored within numerical errors.
While $<\sigma_0>,<\sigma_8>$ have numerical errors
$\Delta<\sigma>/<\sigma>=0.1\%$, the error on the  $<\bar q
q>$ condensate is $\frac{\Delta<\bar qq>}{<\bar qq>}=4\%$.
Systematic errors are attached to the current quark masses. These errors
give the
main contribution to $\frac{\Delta<\bar qq>}{<\bar q q>}=25\%$. An exact
restoration cannot be expected, since the symmetry is explicitly broken in
the
Lagrangian. When we derive the quark condensates from Eq.~(\ref{eq:3}), we
treat the
quark masses and the external fields as temperature independent up to the
transition
region. This assumption may not be justified. The corresponding error is
unknown.

In our lowest order calculation of the effective potential  we cannot
distinguish between pole- and screening masses. By `masses' we mean the
effective
masses $X^2_Q$ entering the arguments of the logarithm according to
Eq.~(\ref{eq:18}),
evaluated at the  physical mass point $\xi_0=<\xi_0>$ and $\xi_8=<\xi_8>$,
cf. e.g. Eqs.~(\ref{eq:10})-(\ref{eq:11}). Thus the temperature
dependence of these masses is determined by the temperature dependence
of the
condensates.
The masses $m_\pi$ and $m_{f'_0}$ become degenerate for temperatures
$T\geq
T_c$ within the errors $m_\pi\approx (139\pm 10$ MeV), $m_{f'_0}\approx
(141\pm 10$
MeV) at $T_c$, when $T_c$ is approached from above. The degeneracy is a
result of the
vanishing light quark condensate. A zero value of $<\bar q q>$ above  $T_c$
implies a
relation between $<\sigma_0>$ and $<\sigma_8>$, which leads to
$m^2_\pi\approx
m^2_{f'_0}$. Below $T_c$ the meson masses stay remarkably constant. The
pion mass
$m_\pi$ and the scalar mass $m_{f'_0}$ change from
$m_\pi=(150\pm 10$ MeV) and $m_{f'_0}=(625\pm 10$ MeV) at $T=T_c$
approaching $T_c$ from below to
$m_\pi=(139\pm 10$ MeV) and $m_{f'_0}=(141\pm 10$ MeV) at $T=T_c$
=161 MeV
approaching $T_c$ from above.

The behaviour of the scalar mass can induce an
abrupt change in temperature  dependent cross sections in contrast to a
smooth
variation in case of a second order transition or a crossover phenomenon.
Temperature
dependent cross sections may be realized in heavy ion collisions, when the
transient
quark gluon plasma cools down to the hadron phase and the hadron phase
evolves until
freeze-out.

The $f'_0$ above $T_c$ can no longer decay into two pions, its width for
$T>T_c$
has a contribution from $f'_0\to2\gamma$ decays with an invariant mass
$m^2(2\gamma)\approx m^2_{\pi^0}$. In the experiment one may see an
anomalous amount
of such $(2\gamma)$ decays, when the system spends a sizeable time in the
phase where
chiral symmetry is restored. In the experiment WA98 at CERN a
measurement of the
number of gammas to the number of charged mesons, i.e. mostly
$\pi^{\pm}$, is planned.
Above $T_c$,  the $\pi-K$ splitting is increased rather than reduced. The
mass difference of $\Delta m=m_K-m_\pi=(400\pm 20$ MeV) below $T_c$
$(T<161$ MeV) is
increased to $\Delta m=(516\pm 2$ MeV) for $T=161\  {\rm MeV} \geq
T_c$. Accordingly
the strange meson contribution to the energy density in this temperature
region is
reduced compared to the low-temperature hadron gas.

In Fig.~3 we give the energy density $\varepsilon/T^4$ and pressure
$p/T^4$ as function
of temperature for a cut-off $\Lambda=$ 1.5 GeV. The gap in  the energy
densities at
$T_c$, which is a measure for the latent heat, is obviously rather small,
about
10\% of $\varepsilon$ at $T_c$.  Sizeable  contributions to $\varepsilon$
come mainly
from 8 degrees of freedom, the pions, the kaons
and the $f_0$ meson.
The small value of the latent heat may be due to the vicinity of a
(hypothetical)
first order boundary in the $(m_K,m_\pi)$-mass diagram. At the location of
this
boundary the meson masses become so large that they wash out the chiral
transition
completely. We plan to check this explanation by investigating the chiral
transition
for unphysical values of the strange quark and light quark masses in the
future.

The existence of quasibound $\sigma(f'_0)$ and  $\pi$ modes may be a good
approximation in the vicinity of $T_c$. Far above $T_c$ the linear sigma
model
certainly fails as an effective model for QCD due to the lack of quark-gluon
degrees
of freedom. Nevertheless it would be interesting to study, at what
temperature
 the full $SU(3)\times SU(3)$ symmetry is restored. At high temperatures
the
effective potential becomes proportional to $\sum_Q X^2 (Q) T^2$, the linear
terms
in the meson masses (Eqs.~(\ref{eq:10})-(\ref{eq:11})) cancel and
temperature tries to
fully restore the broken symmetry.

Finally we remark that our value for
$T_c$ is rather close to the lower limit of the Hagedorn temperature $T_H
(T_H\sim160$~MeV) \cite{hag}. This may not be entirely accidental. In our
$1/N$-expansion $N$ means a large   number of  flavours, since
\begin{equation}\label{eq:31}
N=2\cdot N^2_f.
\end{equation}
In order to keep QCD an asymptotically free theory also  the number of
colours $N_c$
has to increase. Correspondingly our approximation is similar to Hagedorn's
description of the hadron gas as a resonance gas.  We expect that corrections
from
subleading terms in our $1/N_f$-expansion will implicitly amount to
corrections also
to the large $N_c$-limit.

\newpage

\section*{Acknowledgments}

We thank Prof. L. McLerran for his comments and criticism on the present
work, in
particular for his remark concerning the analogy to Hagedorn's theory.

\section*{Figure Captions}
\begin{description}
\item[Figure 1] Normalized light quark condensate
$\displaystyle\frac{\left<\bar{q}q\right>}{\left<\bar{q}q
\right>_{T=0}}$ vs temperature.
\item[Figure 2] Normalized strange quark condensate
$\displaystyle\frac{\left<\bar{s}s\right>}{\left<\bar{s}s
\right>_{T=0}}$ vs temperature.
\item[Figure 3] Normalized energy density
$\displaystyle\frac{\varepsilon}{T^{4}}$ and pressure
$\displaystyle\frac{p}{T^{4}}$
vs temperature. The decrease of these quantities above $T\approx200$~MeV
as a function
of $T$ indicates the breakdown of our approximation scheme.
\end{description}

\end{document}